\documentclass[11pt]{article}

\usepackage{times}
\usepackage{a4wide}
\usepackage{amsmath}
\usepackage{amsfonts}
\usepackage{graphicx}
\usepackage{arydshln}

\newcommand{\ints}{{\mathbb Z}}
\newcommand{\Prob}{{\mathbb P}}

\title{	Technical Report 12:02\\
Statistics Group, School of Mathematics\\
	University of Bristol, UK\\
	Simulation Study Comparing Two Tests of Second-order Stationarity and Confidence Intervals
	for Localized Autocovariance}
	
\author{G. P. Nason}

\date{July 4th 2012}

\begin{document}
\maketitle

\begin{abstract}
This report compares two tests of second-order stationarity through simulation.
It also provides several examples of localised autocovariances and their approximate
confidence intervals on different real and simulated data sets. An empirical verification
of an asymptotic Gaussianity result is also demonstrated. The commands use to
produce figures in a companion paper are also described.
\end{abstract}

\section{Introduction}
This report is a companion to the article Nason (2013).

This technical report contains two sets of simulation studies. The first set concerns the performance
of two different tests of second-order stationarity described in Section~\ref{sec:tos}.
The second set examines confidence intervals
for a particular form of localized autocovariance in Section~\ref{sec:la} and also provides an
empirical justification of approximate normality of a statistic defined in the companion paper.
Finally, Section~\ref{sec:rr} provides R commands for the {\tt locits} package (which implements
the new ideas in the companion paper) showing how the figures were produced in the
companion paper.

In the following we refer to the Priestley-Subba Rao (1969) test as PSR and the test
introduced in the companion paper, the Haar wavelet on wavelet 
spectrum test, as HWTOS. 

Further explanation appears in the sections below, however the main conclusions of this report concerning the
tests are:
\begin{enumerate}

\item the HWTOS tends to be more conservative than PSR, but not always;

\item for light-tailed noise the HWTOS's empirical size tends to be less than its
	nominal size, whereas the PSR tests tend to be slightly above;
	
\item for heavy-tailed data the PSR test does not perform well, giving many false
	positives by rejecting the null hypothesis erroneously about 60\% of the time
	for all models we tried. By comparison the HWTOS tests have much
	better empirical size values, although greater than their nominal size.
	
\item The PSR test has problems (empirical size of about 12\%) for an AR$(1)$ process
	with AR parameter of 0.9. Similarly, the HWTOS tests have problems (empirical size
	of about 20\%) for an AR$(1)$ process with parameter $-0.9$.
	
\item Sometimes PSR is much more powerful than HWTOS and vice versa, sometimes both
	tests do not have much power to detect particular nonstationary alternatives and sometimes
	both have excellent power. As expected power depends on the alternative and the sample
	size. 
\end{enumerate}

\section{Tests of Stationarity}
\label{sec:tos}
The two tests are the Priestley-Subba Rao test of stationarity implemented by the
{\tt stationarity} function in the {\tt fractal} package in {\tt R} and (ii) the test of stationarity
obtained by examining the Haar wavelet coefficients of an evolutionary wavelet spectrum
estimate modelled by locally stationary wavelet processes, implemented in the
{\tt hwtos2} function and introduced by the companion paper. We augment the HWTOS test with two
forms of multiple comparison control: FDR and Bonferonni and, as such, refer to two tests.

For both cases we are concerned with a given time series
$x_t$ for $t=1, \ldots, T$ which is a realisation from a stochastic process
$\{ X_t \}_{t \in \ints}$. The null hypothesis that we wish to test concerning $X_t$ is $H_0:$ the process
is stationary versus $H_A:$ it is not! 

\subsection{Evaluation of statistical size}
In this section we provide simulation evidence of the empirical statistical size of the various stationarity tests.
Recall that the size of a hypothesis test is $\gamma = \Prob ( \text{Reject $H_0$} | H_0)$.
To empirically evaluate this we first set a nominal
size in our test procedure $\gamma =0.05$.
(Note that $\alpha$ is often used as the symbol for size but here we used $\gamma$ and $\alpha$ is reserved
for the parameter of an AR$(1)$ model.)
Then we run the testing procedure on $N$ realisations from  a known
stationary model and count how many times the test rejects the null hypothesis, $R$. The empirical size of
the test is then given by $R/N$. A test is conservative if $R/N < \gamma$ and conservative tests are
generally thought of as desirable and not given to reporting false positives.

The various stationary models that we have considered are:

\begin{description}

\item [S1] iid standard normal;

\item [S2] AR$(1)$ model with AR parameter of $0.9$ with standard normal innovations;

\item [S3] As S2 but with AR parameter of $-0.9$;

\item [S4] MA$(1)$ model with parameter of $0.8$;

\item [S5] As S4 but with parameter of $-0.8$.

\item [S6] ARMA$(1,0,2)$ with AR parameter of -0.4, and MA parameters of $(-0.8, 0.4)$.

\item [S7] AR$(2)$ with AR parameters of $\alpha_1 = 1.385929$ and $\alpha_2 -0.9604$.
	The roots associated with the auxiliary equation, (see Chatfield, {\em The Analysis of Time Series}, book)
	are $\beta_1 = \bar{\beta_2} = 0.98 e^{i \pi/4}$. This process is stationary, but close to the `unit root': a `rough' stochastic
	process with spectral peak near $\pi/4$.

\end{description}

The empirical size values are given in Table~\ref{tab:size}, these are all computed over $N=1000$ realisations
with a sample size of $T=512$ for the realization.

\begin{table}
\centering
\caption{Empirical size estimates (\%) for stationary models with nominal size of 5\% \label{tab:size}}
\begin{tabular}{r|rrr}\hline
Model & PSR & HWTOS (Bon) & HWTOS (FDR)\\\hline
S1 &  5.6 & 4.3 & 4.3\\
S2 & 12.4 & 4.0 & 4.7\\
S3 & 6.2 & 20.3 & 20.5\\
S4 & 6.0 & 3.4 & 3.8\\
S5 & 6.5 & 0.7 & 0.7\\
S6 & 7.5 & 0.1 & 0.1\\
S7 & 23.9& 7.3 & 7.4\\\hline
\end{tabular}
\end{table}

\subsection{Further investigation of S3}
\begin{figure}
\centering
\resizebox{\textwidth}{!}{\includegraphics{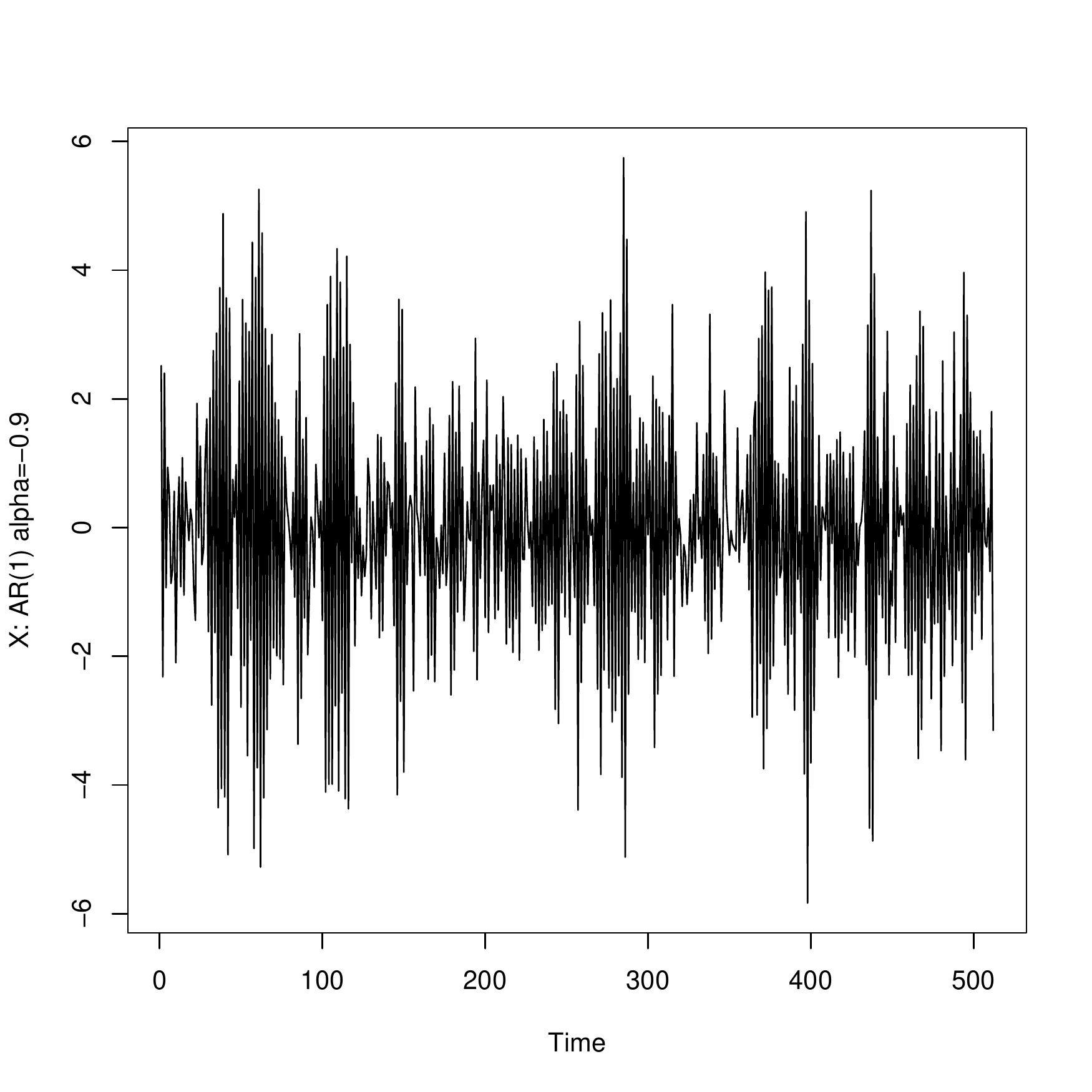}}
\caption{Realization of stationary AR$(1)$ process with AR parameter of -0.9\label{fig:ar1}}
\end{figure}

In the previous section the empirical size values from the S3 model were about 20\% for the Haar wavelet based test
of stationarity, even the Priestley-Subba Rao test has a  estimated power at 6.2\%, a full 1.1\% above the nominal.
What is going on? Figure~\ref{fig:ar1} shows a realisation from the AR$(1)$ process with parameter of $\alpha =-0.9$
and this indicates what might be happening. The realisation in Figure~\ref{fig:ar1} clearly shows a kind of `volatility clustering',
which is reminiscent with what can happen with GARCH processes although, in this case, unlike GARCH, the regular
autocorrelation is non-zero. 
The Haar wavelet test appears to be reading the volatility clustering as non-stationarity. However, one might well ask
what a human would say if confronted with just Figure~\ref{fig:ar1} the local variance appears to be changing quite
markedly and so we would submit that there is a strong chance that a human observer might regard the series
at non-stationary or at least be suspicious.

Another thing to note is that about 17\% of the occasions that the series was deemed non-stationary (out of the 20\%) only
{\em one} Haar wavelet coefficient out the 186 tested simultaneously each time was assessed to be significantly different
from zero. So, in roughly 17\% of cases one would probably practically say that the series is, in fact,
very close to stationarity as a more markedly non-stationary
series would result in many more Haar coefficients to be assessed as significant. Note, that for this AR process only 3\%
had two significant coefficients and only 0.2\% had three significant coefficients and there were never more than three.
Hence, the empirical size values are a bit of blunt instrument. In practice, we believe users will be more interested
in the numbers of significant Haar coefficients, where they are and maybe not pay too much attention to situations where very few
coefficients are significant. This remains to be seen.

If $\alpha = -0.8$ in an AR$(1)$ model
then the estimated size values are 5.0\% for PSR, 3.7\% for HWTOS(Bon) and 3.9\% for HWTOS(FDR),
so for this less extreme value of the AR parameter the HWTOS again becomes more conservative than PSR and we have the
desirable characteristic of empirical size smaller than nominal size.

If $\alpha= -0.99$ then the estimated size values are 7.3\% for PSR and about 80\% for the Haar wavelet tests. So, the empirical
size for this extreme case for the Haar wavelet test is not good, but on further examination the largest number of significant
Haar coefficients out the 186 tested each time is never more than 4. Indeed, about 34\% involve one significant coefficient,
33\% involve two, 12\% involve three and only about 0.3\% involve four. So, pragmatically, even for the extreme AR case
the Haar test indicates that the series is very close to stationarity.

\subsection{Empirical size results for heavy tailed data}
We still would like our test to perform well with heavy-tailed data and for it still to assess
stationarity accurately. Data with heavy tails tends to cause the number of false positives to
increase (i.e.\ detect non-stationarity when the series is, in fact, stationary). Hence,
we repeat our empirical size analysis, but this time replace all the normal
variates and/or innovations by samples from the (a) double-exponential distribution and
(b)  \mbox{Student's-$t$} distribution
on four degrees of freedom. (This can be achieved in {\tt arima.sim} in R through its
{\tt rand.gen} argument). We denote the new models by SHD1--SHD7 and SHT1--SHT7 respectively which are the same
as S1--S7 but with heavy-tailed innovations. Note that the SHD models fall into the class of processes that
adhere to the distributional constraints indicated in the companion paper (Assumption 3), but the SHT do not (and hence the
latter are an extreme test of the methods).

\begin{table}
\centering
\caption{Simulated size estimates (\%) for stationary exponential-tailed models with nominal size of 5\% \label{tab:sizeht}}
\begin{tabular}{r|rrr}\hline
Model & PSR & HWTOS (Bon) & HWTOS (FDR)\\\hline
SHD1 &  43.8 & 7.3 & 7.9\\
SHD2 & 48.9 & 5.8 & 7.0\\
SHD3 & 40.6 & 20.5 & 20.8\\
SHD4 & 44.5 & 7.1 & 7.8\\
SHD5 & 46.8 & 15 & 19\\
SHD6 & 45.1 & 11 & 12\\
SHD7 & 57.6 & 10.6 & 11.4\\\hline
\end{tabular}
\end{table}

\begin{table}
\centering
\caption{Simulated size estimates (\%) for stationary \mbox{$t$-distribution}  models with nominal size of 5\% \label{tab:sizeht}}
\begin{tabular}{r|rrr}\hline
Model & PSR & HWTOS (Bon) & HWTOS (FDR)\\\hline
SHT1 &  60.3 & 15.1 & 16.7\\
SHT2 & 64.9 & 9.8 & 11.3\\
SHT3 & 63.9 & 28.3 & 28.7\\
SHT4 & 62.7 & 15.8 & 18.0\\
SHT5 & 63.0 & 7.1 & 8.1\\
SHT6 & 63.0 & 6.6  & 6.8\\
SHT7 & 69.0 & 14.1 & 15.4\\\hline
\end{tabular}
\end{table}

The results in Table~\ref{tab:sizeht} are interesting: both the HWTOS and PSR test's empirical
size is greater than the nominal size of 5\%. However, the PSR is much less robust than the HWTOS
options.  One could not recommend PSR for heavy-tailed observations, whereas HWTOS 
is a more realistic proposition (because if even if the series was stationary PSR will falsely reject
the null hypothesis approximately 60\% of the time).

\subsection{Power Simulations}
\begin{figure}
\centering
\resizebox{\textwidth}{!}{\includegraphics{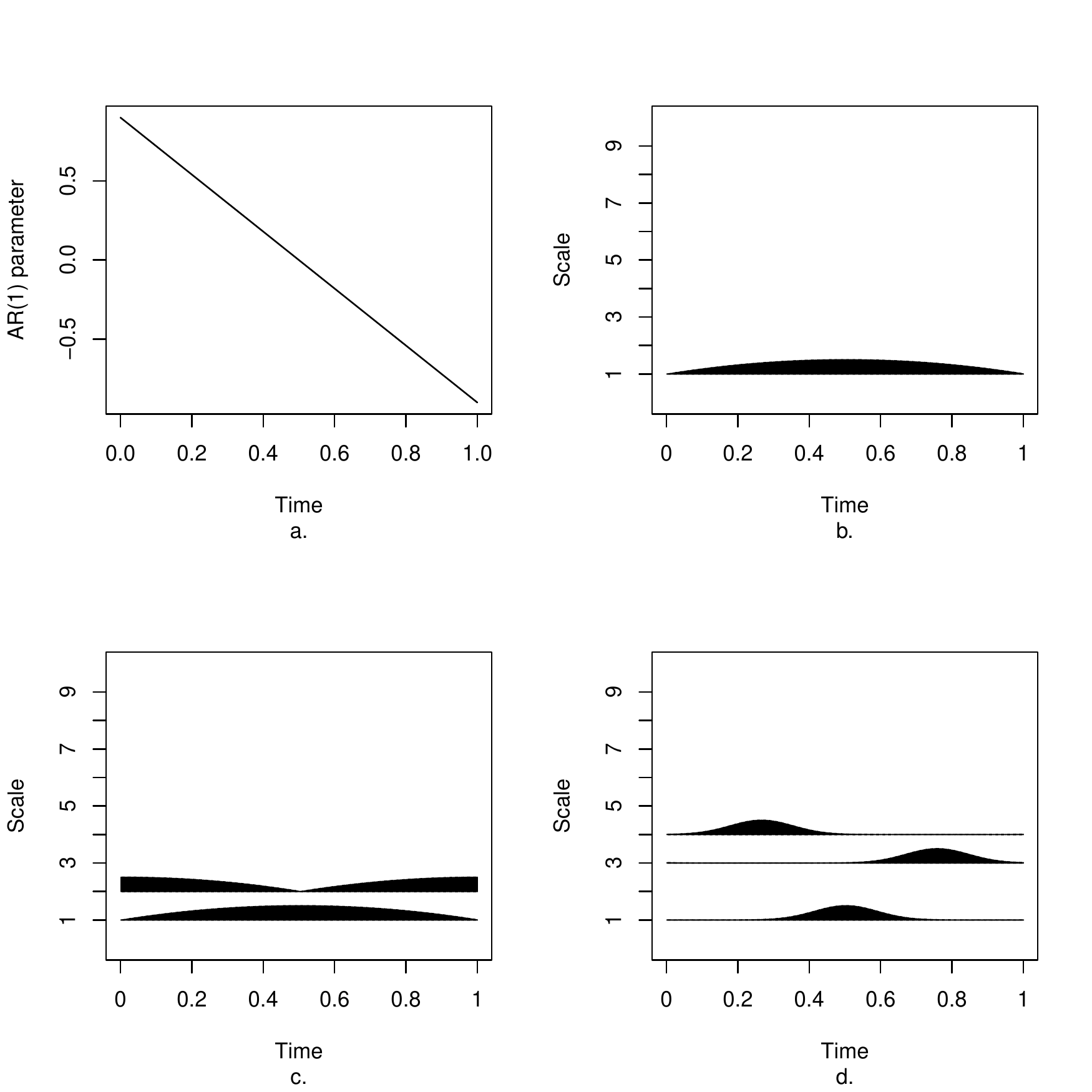}}
\caption{\label{fig:specs} {\em a.} AR$(1)$ parameter as it varies over time for model P1. {\em b.-- d.} evolutionary wavelet
	spectra underlying processes P2, P3 and P4 respectively.}
\end{figure}

\begin{figure}
\centering
\resizebox{\textwidth}{!}{\includegraphics{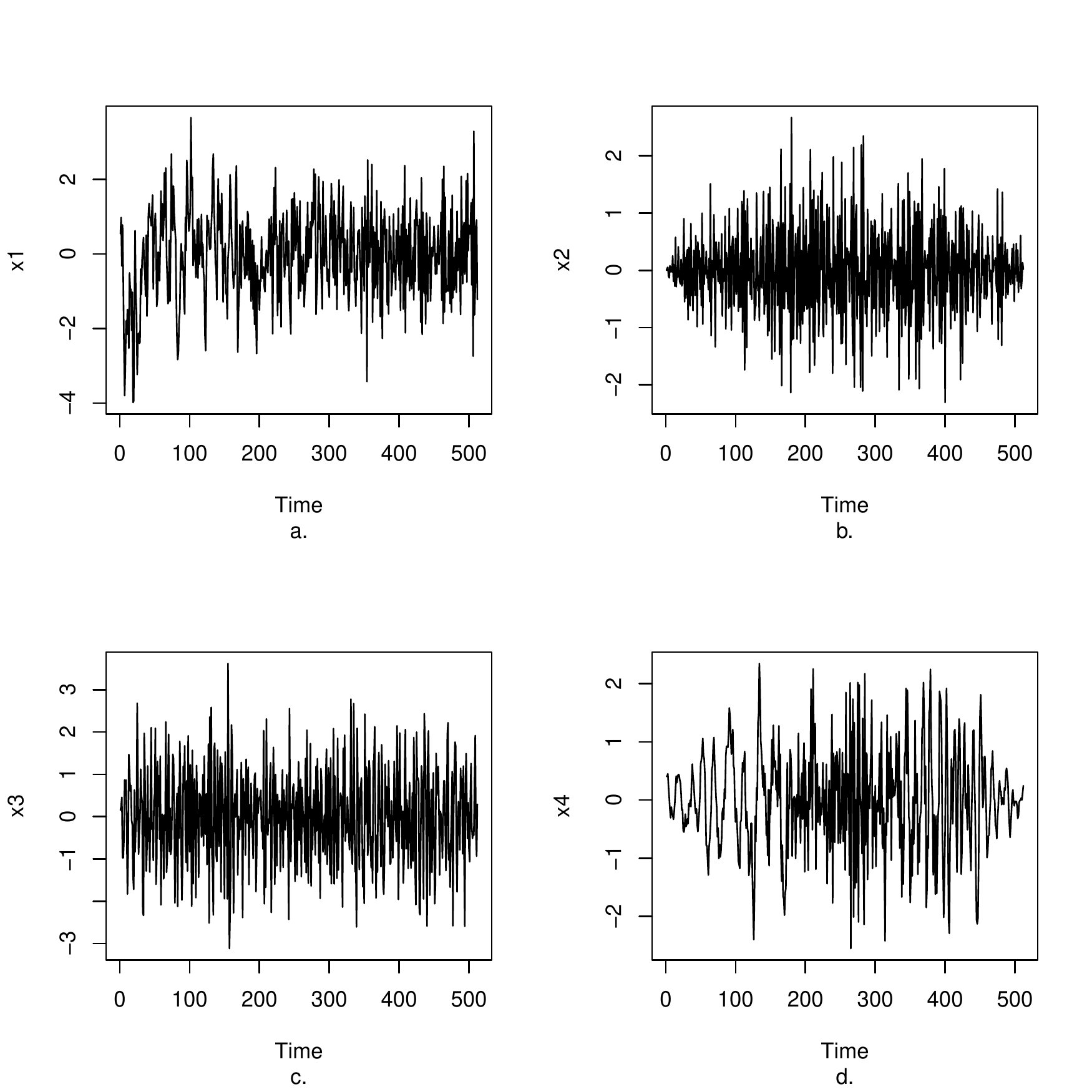}}
\caption{\label{fig:realiz} {\em a.--d.} A single realisation from each of P1 thru P4 respectively.}
\end{figure}

To explore statistical power we need to create nonstationary processes and then count the number
of times each test decides a realisation is not stationary over multiple realisations. The models we choose are:

\begin{description}

\item [P1] Time-varying AR model $X_t = \alpha_t X_{t-1} + \epsilon_t$ with iid standard normal innovations and
	the AR parameter evolving linearly from 0.9 to -0.9 over the 512 observations.
	
\item [P2] A LSW process based on Haar wavelets with spectrum $S_j(z) = 0$ for $j>1$ and $S_1(z) = \tfrac{1}{4} - (z-\tfrac{1}{2})^2$
for $z \in (0,1)$. This process is, of course, a time-varying moving average process.

\item [P3] A LSW process based on Haar wavelets with spectrum $S_j(z) = 0$ for $j>2$ and $S_1(z)$ as for P2 and
	$S_2(z) = S_1(z + \tfrac{1}{2})$ using periodic boundaries (for the construction of the spectrum only).
	
\item [P4] A LSW process based on Haar wavelets with spectrum $S_j(z) = 0$ for $j=2, j >4$ and
	$S_1(z) = \exp \{ -4(z-\tfrac{1}{2})^2 \}$, $S_3(z) = S_1(z - \tfrac{1}{4})$, $S_4(z) = S_1(z + \tfrac{1}{4})$ again
	assuming periodic boundaries.
	
\end{description}
The parameter profile for P1 and the spectra are associated with P2-P4 are plotted in Figure~\ref{fig:specs}.
A realisation from each of P1--P4 is shown in Figure~\ref{fig:realiz}.

Empirical power values for the PSR and Haar wavelet test can be found in Table~\ref{tab:power}. Once again, these
are computed over $N=1000$ realisations and the nominal size of the test was 5\%.
\begin{table}
\centering
\caption{Empirical power estimates (\%) for models P1--P4 with nominal size of 5\% \label{tab:power}}
\begin{tabular}{r|rrr}\hline
Model & PSR & HWTOS (Bon) & HWTOS (FDR)\\\hline
P1 &  37.2 & 99.7 & 99.9\\
P2 & 100 & 17.3 & 19.2\\
P3 & 44.3 & 1.3 & 1.3\\
P4 & 100& 94.8 & 97.8\\\hline
\end{tabular}
\end{table}
The simulation results for power paint an interesting picture. Sometimes the HWTOS tests are
good and the PSR is not (P1), sometimes PSR is good and HWTOS is not (P2), sometimes both
are not that good (P3) and sometimes both are very good (P4).

Sample size is an important determining factor in power: increasing sample size should increase
power of detection of alternatives. For example, with P2 which has a sample
size of $T=512$ the HWTOS tests have fairly low powers of 17.3\%/19.2\% respectively.
For $T=1024$ the tests are more powerful having powers of 70.7\%/75.2\% and for
$T=2048$ the empirical powers are both 100\%.

\section{Localized Autocovariance}
\label{sec:la}

\subsection{Examples of Localized Autocovariance and CI computation}
In this section we show some examples of the localised autocovariance and 95\% confidence intervals.
We start with four models and draw one realisation of length $T=512$ from each. The four models are:
\begin{description}

\item [AC1] IID standard normal random variables (stationary).

\item [AC2] Stationary AR$(1)$ process with AR parameter of 0.8.

\item [AC3] Time-varying AR$(1)$ process with AR parameter ranging from 0.9 to -0.9 (same model as 
P1, above).

\item [AC4] Time-varying MA$(1)$ process with MA parameter ranging from 1 to -1 over length of the series:
	$X_t = Z_t + \beta_t Z_{t-1}$.

\end{description}
All innovations are standard normal.

\begin{figure}
\centering
\resizebox{\textwidth}{!}{\includegraphics{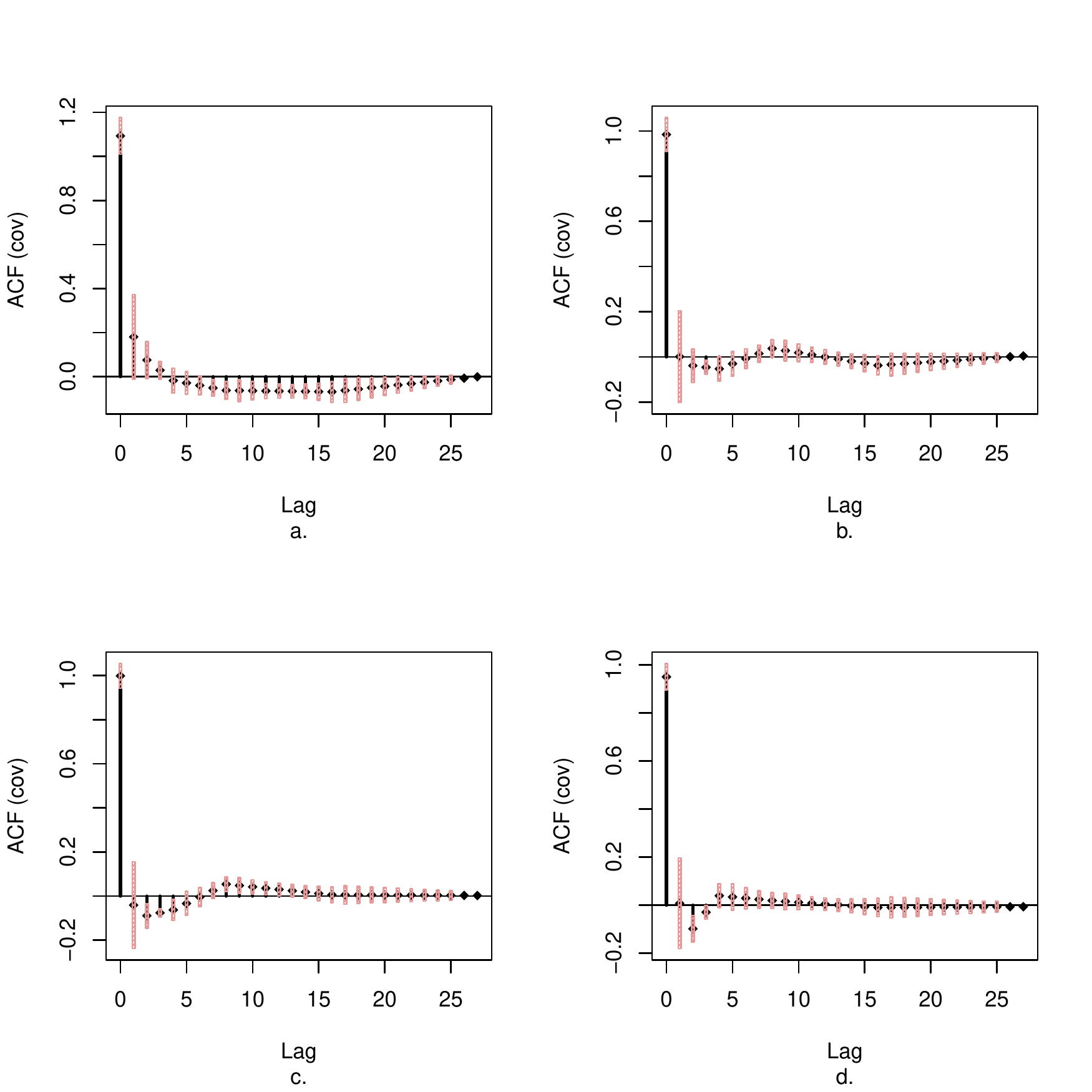}}
\caption{\label{fig:acf1} Localized autocovariance estimates from a single realisation from stationary model AC1.
	Plots {\em a.} to {\em d.} correspond to localizing at (rescaled) times of $100/512, 200/512, 300/512, 400/512$ respectively.}
\end{figure}

\begin{figure}
\centering
\resizebox{\textwidth}{!}{\includegraphics{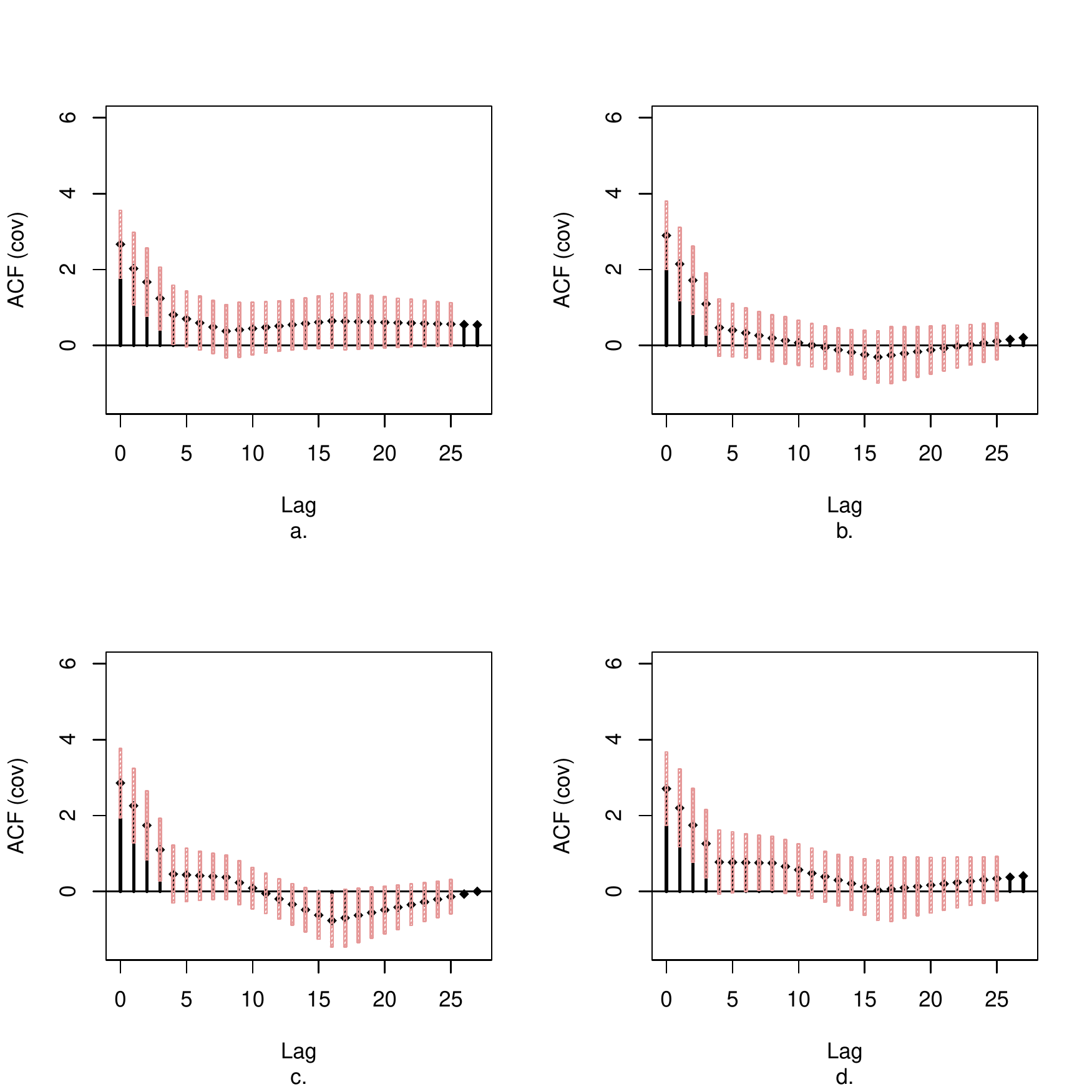}}
\caption{\label{fig:acf2} Localized autocovariance estimates from a single realisation from stationary model AC2.
	Plots {\em a.} to {\em d.} correspond to localizing at (rescaled) times of $100/512, 200/512, 300/512, 400/512$ respectively.}
\end{figure}

\begin{figure}
\centering
\resizebox{\textwidth}{!}{\includegraphics{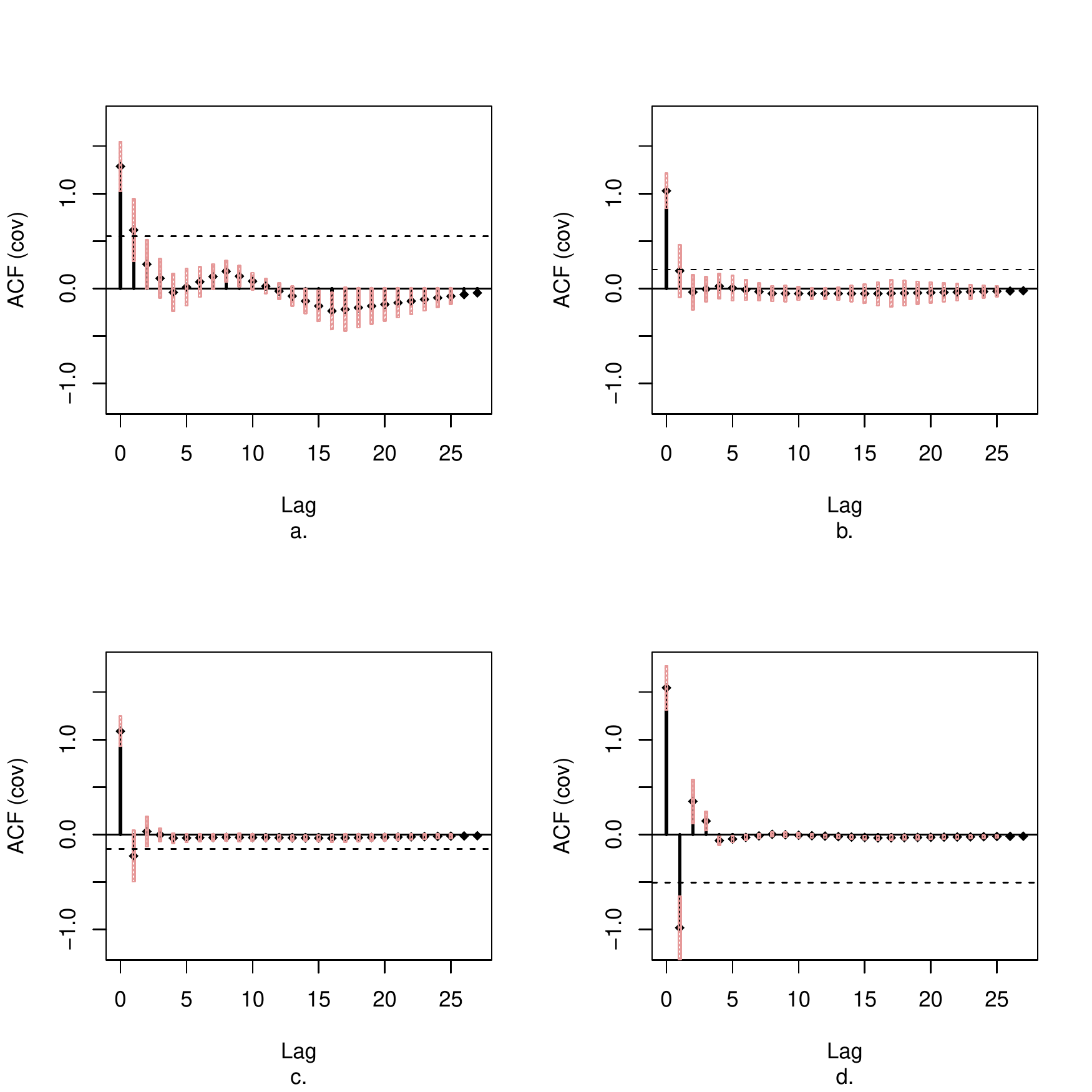}}
\caption{\label{fig:acf3} Localized autocovariance estimates from a single realisation from non-stationary model AC3.
	Plots {\em a.} to {\em d.} correspond to localizing at (rescaled) times of $100/512, 200/512, 300/512, 400/512$ respectively.
	The horizontal dashed line in each figure corresponds to the true value of the AR parameter at that time point}
\end{figure}

\begin{figure}
\centering
\resizebox{\textwidth}{!}{\includegraphics{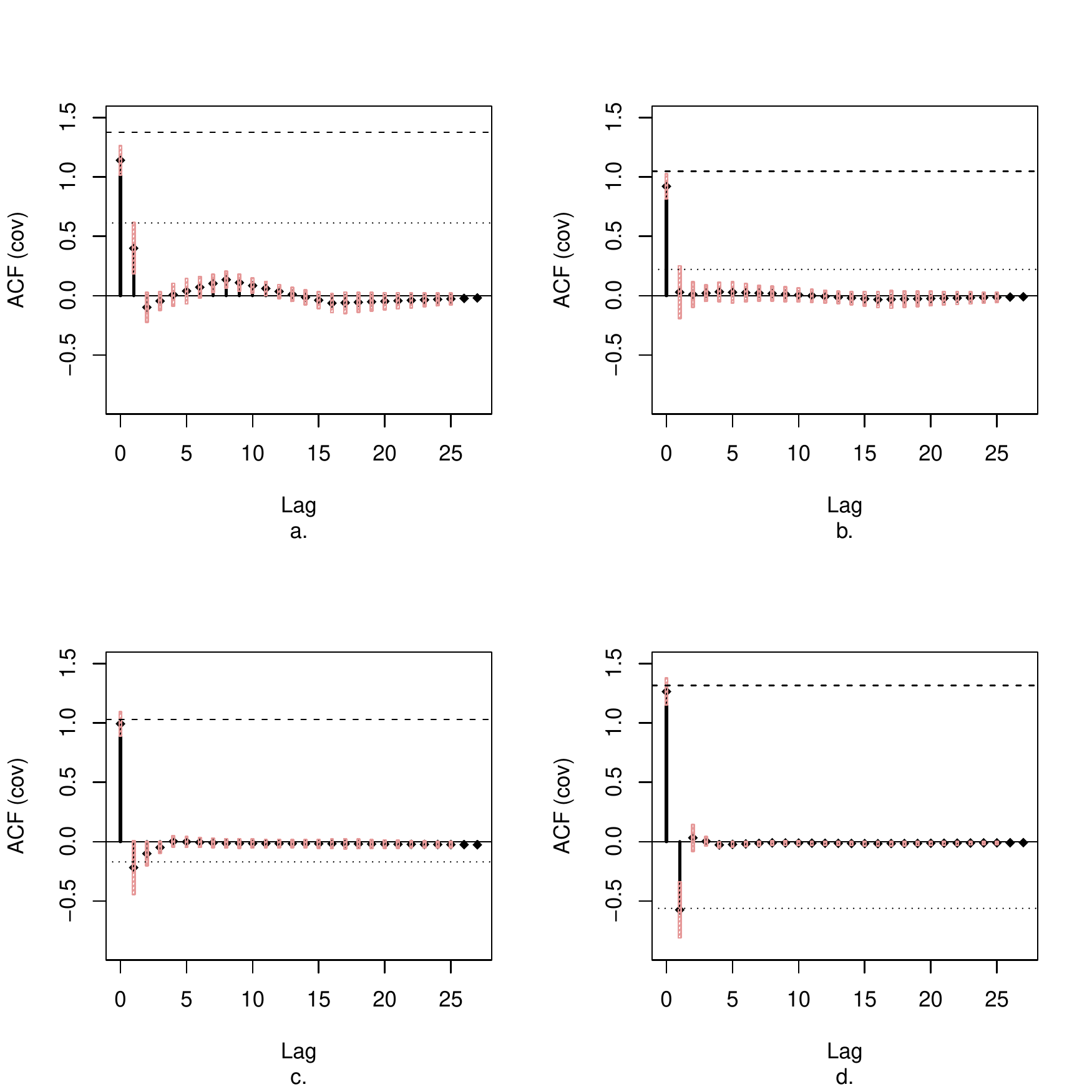}}
\caption{\label{fig:acf4} Localized autocovariance estimates from a single realisation from non-stationary model AC4.
	Plots {\em a.} to {\em d.} correspond to localizing at (rescaled) times of $100/512, 200/512, 300/512, 400/512$ respectively.
	The horizontal dashed line corresponds to the true variance (lag 0 acf) of the process at that time point. The horizontal dotted
	line corresponds to the true values of the lag-1 coefficient at that point.}
\end{figure}

We then compute the
localized autocovariance estimate $\hat{c} (z, \tau)$ for $z=100/512$, $200/512$, $300/512$, $400/512$ in 
rescaled time. Hence, for each model we display a Figure with four localised autocovariance estimates:
these are in Figures~\ref{fig:acf1}, \ref{fig:acf2}, \ref{fig:acf3} and~\ref{fig:acf4}.

Broadly speaking, Figure~\ref{fig:acf1} shows the correct kind of autocovariance plot for each local time point in that
it is very near to one at lag zero, and very near zero for all other lags. The actual $\hat{c}(z, 0)$ value is very close to
1 in plots b.--d.\, a bit further in plot a.\ but all of the 95\% CI cover 1. The acfs at the other lags for the other time
points are all very close to zero: the lag 1 CI always covers zero. However, it should be said that for a few of the acfs at some
lags (notably plot a.) are very small, but their CIs do not cover zero.

Each of the plots in Figure~\ref{fig:acf2} look pretty reasonable to each arise from an AR$(1)$ model.

Figure~\ref{fig:acf3} shows localised acf plots for a time-varying AR$(1)$ model where the AR parameter varies smoothly
from 0.9 to -0.9 over the period of the series. In each of the autocovariance plots the horizontal
dashed line shows the value of the AR parameter at the particular time point. So, for plot a.\ in Figure~\ref{fig:acf3} the
true value of the AR parameter is approximately 0.551 and the estimated $\hat{c}(100/512, 1)$ is very close
to that value and is easily covered by the associated CI.  Plots {\em b. and c.} show similar things (albeit with different
AR parameters). In plot {\em d.} the true value of the parameter is about -0.505, but the estimate is nearly -1.0 and the
CI does not cover the true value.

Figure~\ref{fig:acf4} shows localised acf plots for a time-varying MA$(1)$ model where the MA parameter varies from 
1 to -1 over the length of the series. The horizontal lines show the value of the true lag 0 (variance) and lag 1 acfs, which
are also changing over the length of the series. The CIs all cover the true values (at lags 0 and 1) apart from plot {\em a.}
where the estimate underestimates the truth. Nearly all of the other lags look good (estimated near zero, and a CI which covers zero)
apart from, again, plot {\em a.} where lags 7,8 and 9 appear to be significantly different from zero, but note, still very small.

\subsection{Asymptotic normality check}
This section describes the results of simulations designed to check the asymptotic normality assumption
for the $\hat{c}(z, \tau)$ quantity. To do this we used models AC2 and AC3 and simulated the value of
$\hat{c}(z,\tau)$ $N=1000$ times evaluated at $z = 200/512$ for different values of $T=512, 1024, 2048$ and $T=4096$.

\begin{figure}
\centering
\resizebox{\textwidth}{!}{\includegraphics{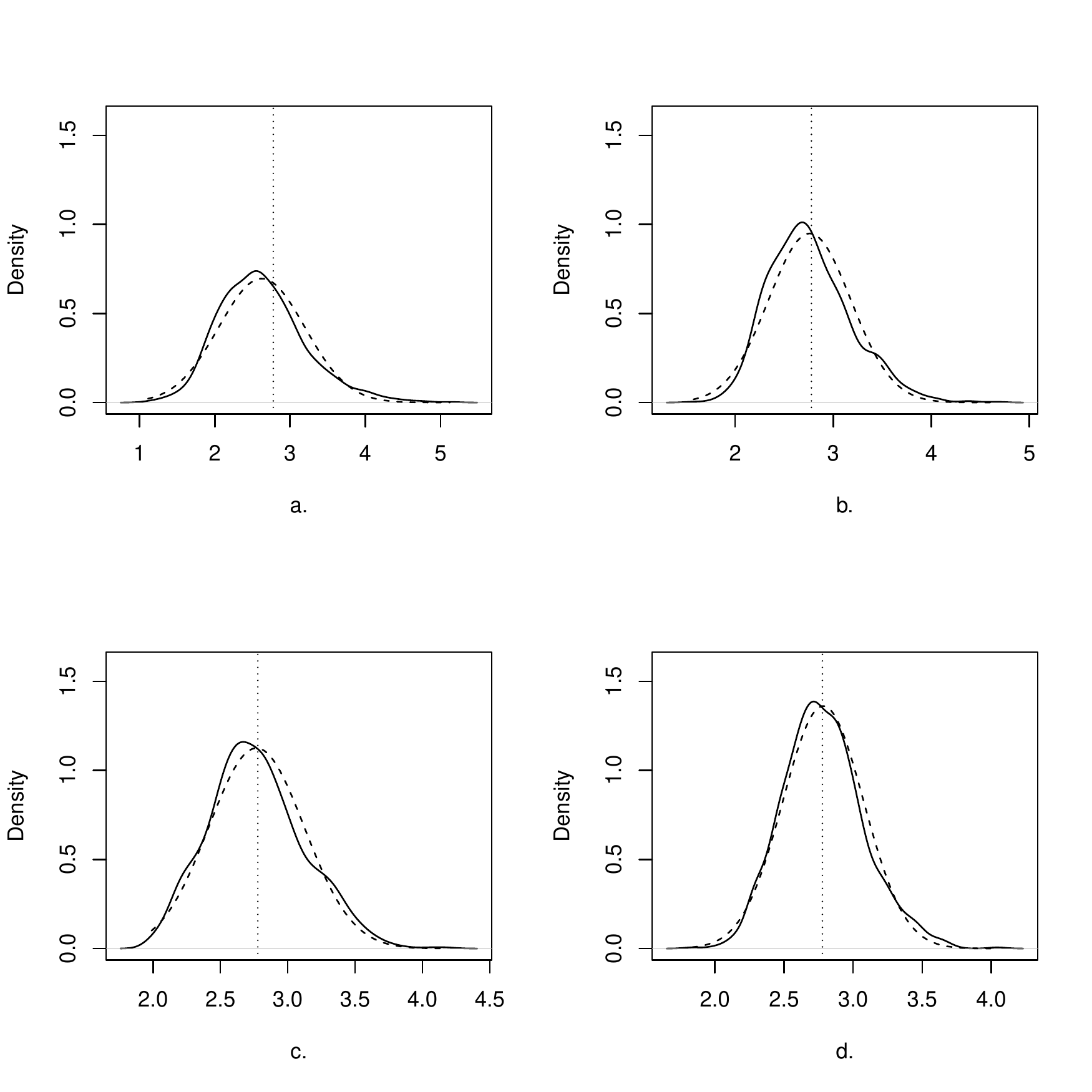}}
\caption{\label{fig:asynAR0} {\em Solid line:} Density estimates of values of $\hat{c}(200/512, 0)$ for
{\em a.} $T=512$, {\em b.} $T=1024$, {\em c.} $T=2048$ and {\em d.} $T=4096$ all
computed from AR$(1)$ model simulations with AR parameter $\alpha=0.8$. {\em Dashed lines:} these
are the normal density functions with mean and variance equal to the sample mean and variance of the
$\hat{c}$ values. {\em Vertical dotted line:} The theoretical value of the parameter.}
\end{figure}

\begin{figure}
\centering
\resizebox{\textwidth}{!}{\includegraphics{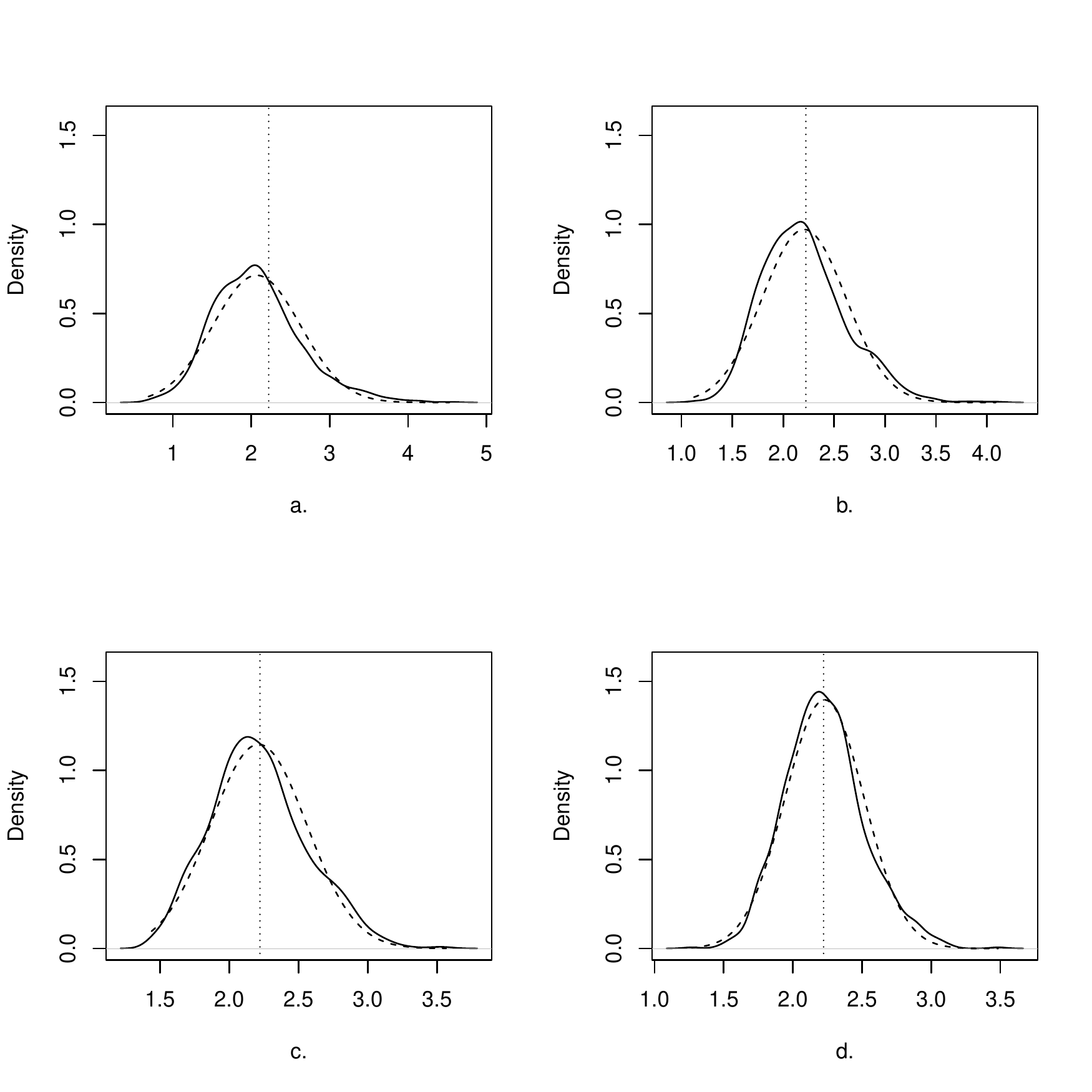}}
\caption{\label{fig:asynAR1} {\em Solid line:} Density estimates of values of $\hat{c}(200/512, 1)$ for
{\em a.} $T=512$, {\em b.} $T=1024$, {\em c.} $T=2048$ and {\em d.} $T=4096$ all
computed from AR$(1)$ model simulations with AR parameter $\alpha=0.8$. {\em Dashed lines:} these
are the normal density functions with mean and variance equal to the sample mean and variance of the
$\hat{c}$ values. {\em Vertical dotted line:} The theoretical value of the parameter.}
\end{figure}

\begin{figure}
\centering
\resizebox{\textwidth}{!}{\includegraphics{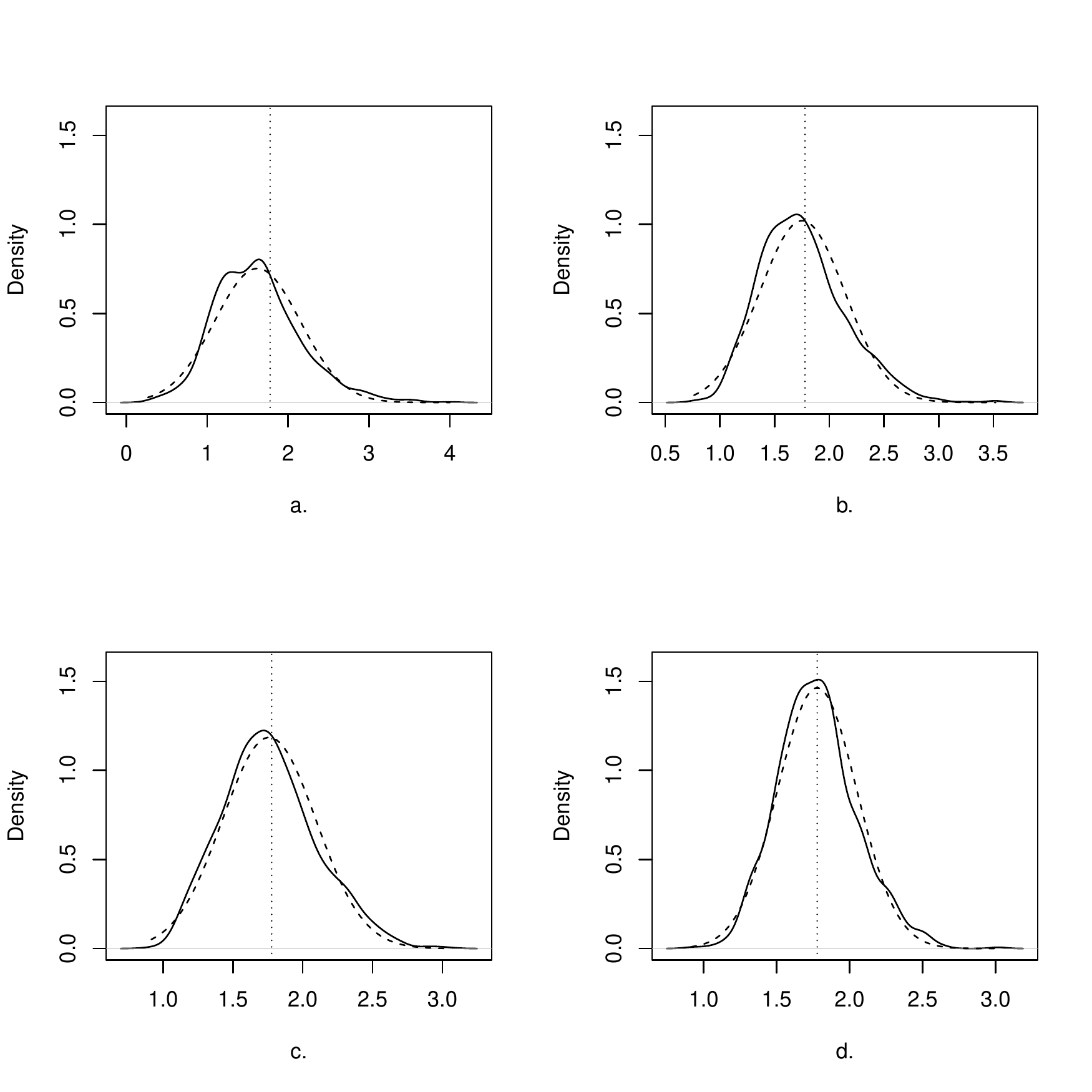}}
\caption{\label{fig:asynAR2} {\em Solid line:} Density estimates of values of $\hat{c}(200/512, 2)$ for
{\em a.} $T=512$, {\em b.} $T=1024$, {\em c.} $T=2048$ and {\em d.} $T=4096$ all
computed from AR$(1)$ model simulations with AR parameter $\alpha=0.8$. {\em Dashed lines:} these
are the normal density functions with mean and variance equal to the sample mean and variance of the
$\hat{c}$ values. {\em Vertical dotted line:} The theoretical value of the parameter.}
\end{figure}

\begin{figure}
\centering
\resizebox{\textwidth}{!}{\includegraphics{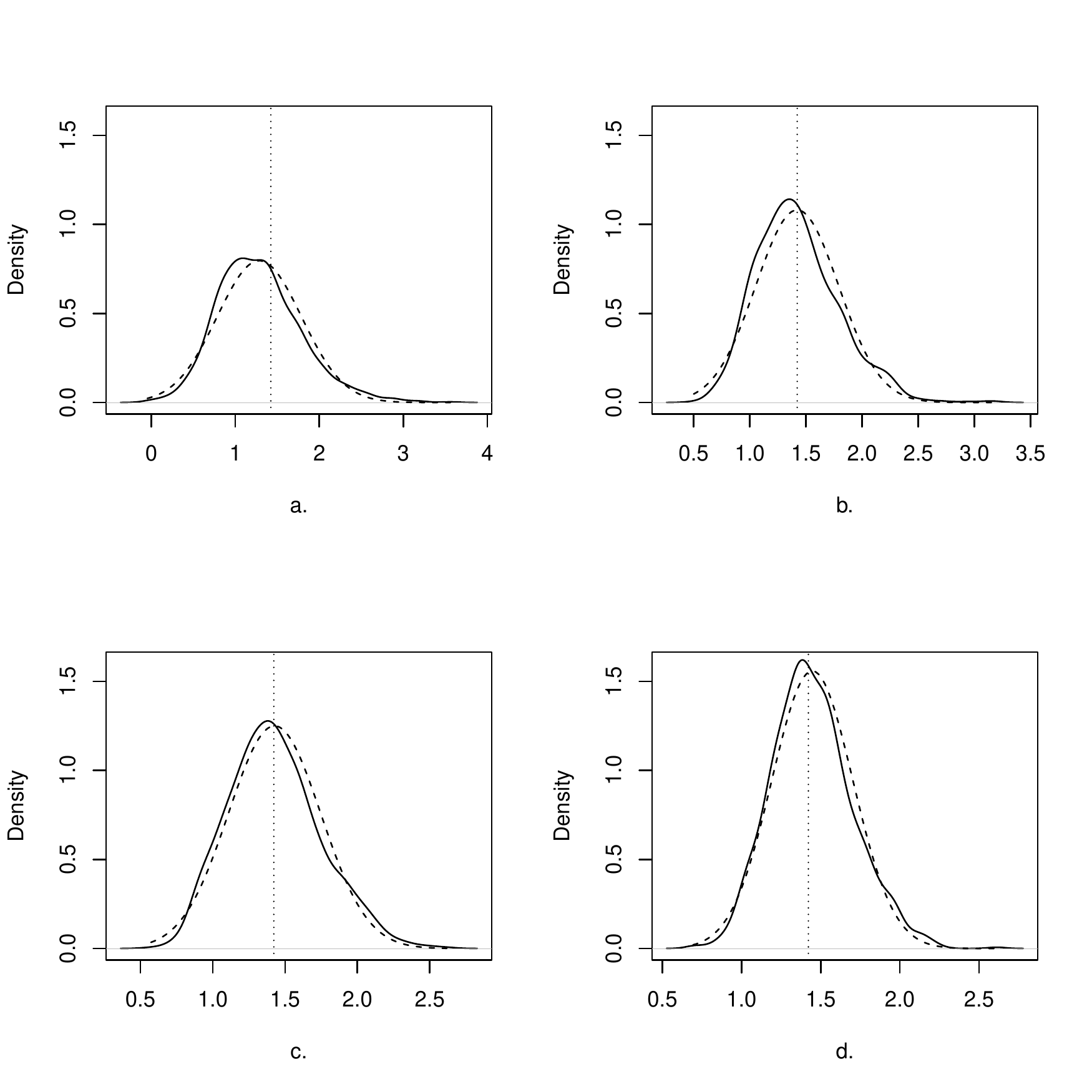}}
\caption{\label{fig:asynAR3} {\em Solid line:} Density estimates of values of $\hat{c}(200/512, 3)$ for
{\em a.} $T=512$, {\em b.} $T=1024$, {\em c.} $T=2048$ and {\em d.} $T=4096$ all
computed from AR$(1)$ model simulations with AR parameter $\alpha=0.8$. {\em Dashed lines:} these
are the normal density functions with mean and variance equal to the sample mean and variance of the
$\hat{c}$ values. {\em Vertical dotted line:} The theoretical value of the parameter.}
\end{figure}

Figures~\ref{fig:asynAR0} to~\ref{fig:asynAR3} show density estimates for the $\hat{c}(200/512, \tau)$ values for
$\tau=0, 1, 2, 3$ and, in each figure, for four different realisation sizes $T=512, 1024, 2048$ and $4096$
with the realisations coming from a stationary AR$(1)$ model with $\alpha=0.8$.
The vertical dotted line in each plot is the theoretical parameter.
Even the plot for $T=512$ in each case does not look too far from normal and would be at least tenable to use its
variance for the basis of a confidence interval.
In each plot one can see  that as the sample size increases
the density estimate of the relevant $\hat{c}$ exhibits far less skew, less kurtosis, less biased and hence
suggests that  the asymptotic normal assumption is tenable. 

\begin{figure}
\centering
\resizebox{\textwidth}{!}{\includegraphics{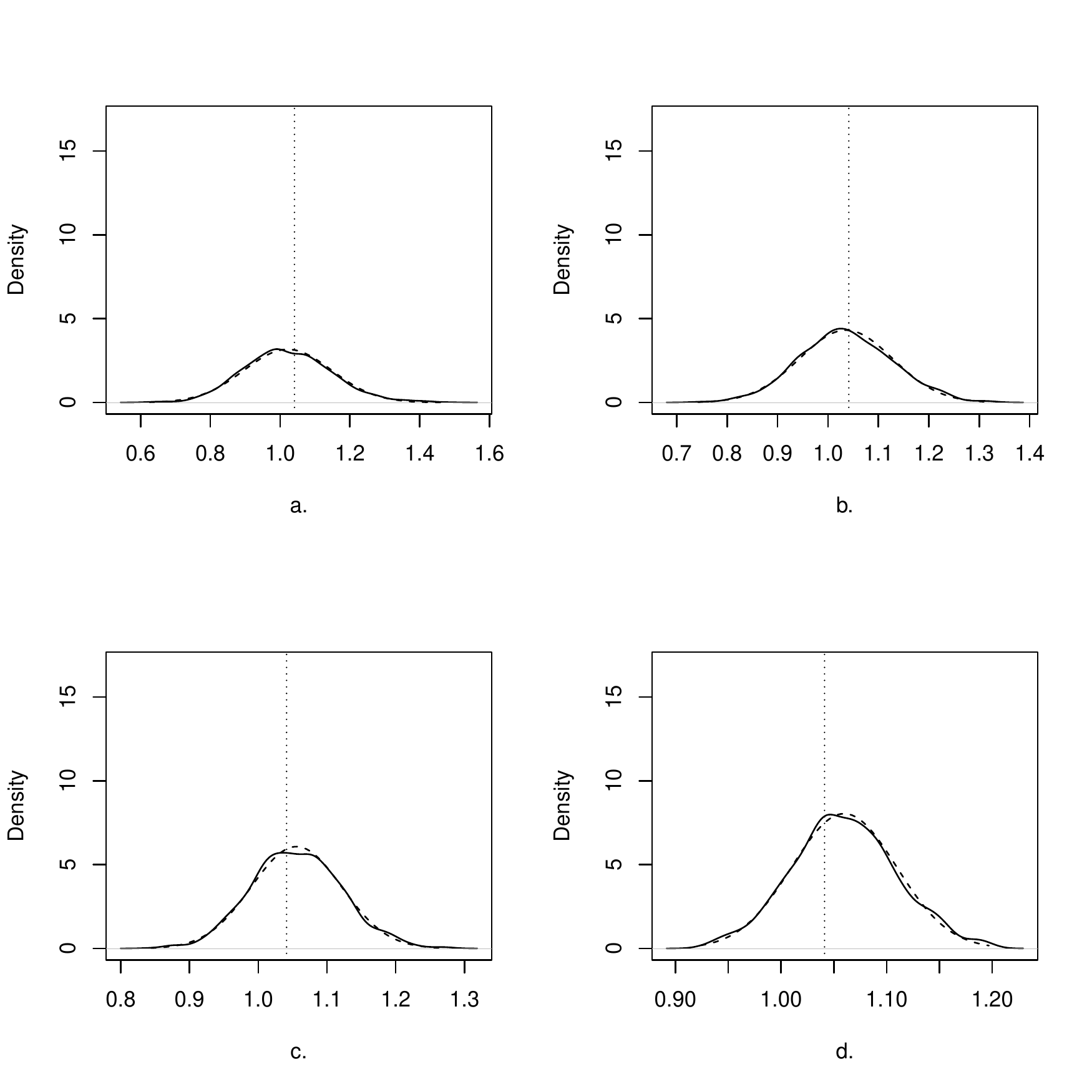}}
\caption{\label{fig:asynTVAR0} {\em Solid line:} Density estimates of values of $\hat{c}(z^{\ast}=200/512, 0)$ for
{\em a.} $T=512$, {\em b.} $T=1024$, {\em c.} $T=2048$ and {\em d.} $T=4096$ all
computed from TVAR$(1)$ model simulations with TVAR parameter $\alpha(z^{\ast})=0.199$. {\em Dashed lines:} these
are the normal density functions with mean and variance equal to the sample mean and variance of the
$\hat{c}$ values. {\em Vertical dotted line:} The theoretical value of the parameter.}
\end{figure}

\begin{figure}
\centering
\resizebox{\textwidth}{!}{\includegraphics{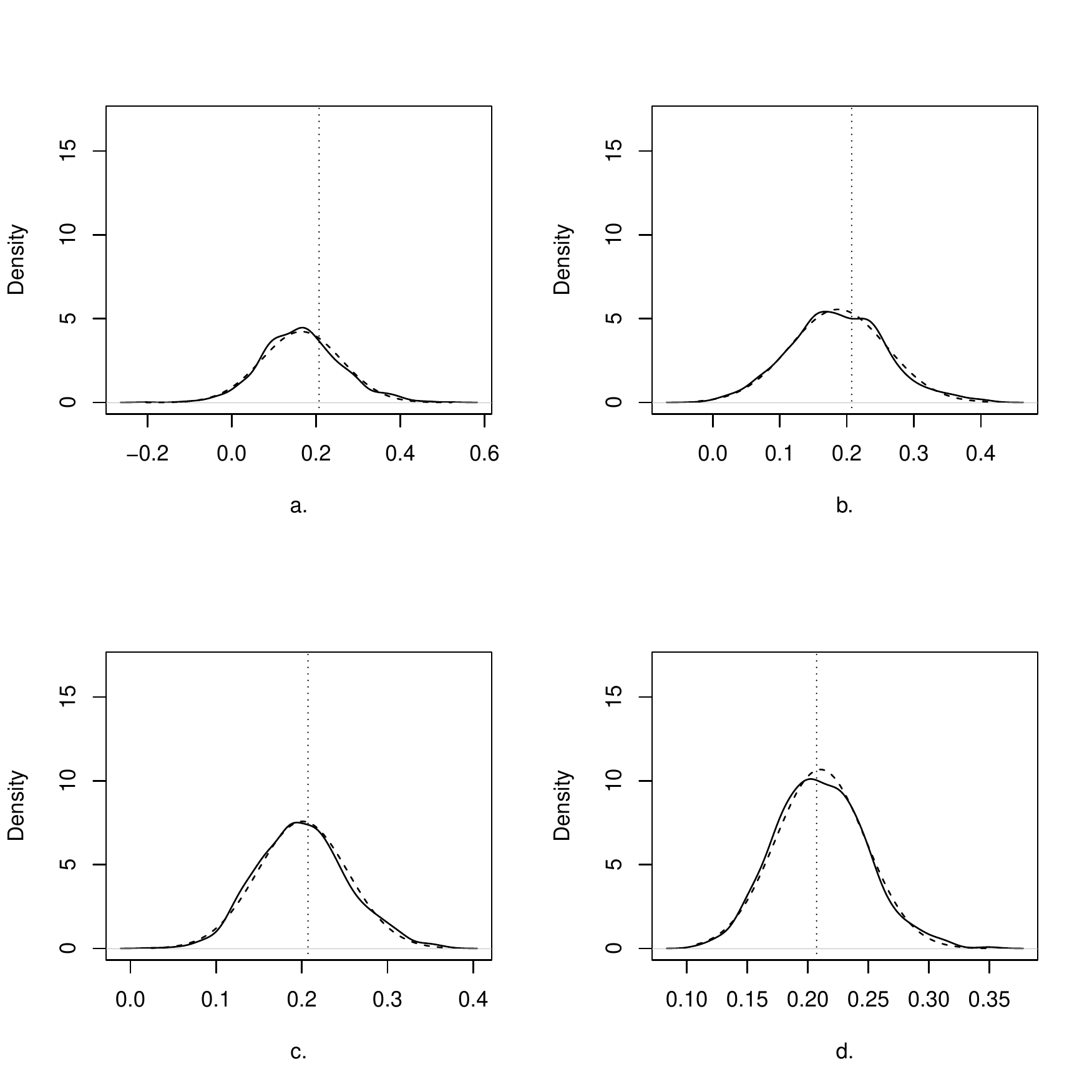}}
\caption{\label{fig:asynTVAR1} {\em Solid line:} Density estimates of values of $\hat{c}(z^{\ast}=200/512, 1)$ for
{\em a.} $T=512$, {\em b.} $T=1024$, {\em c.} $T=2048$ and {\em d.} $T=4096$ all
computed from TVAR$(1)$ model simulations with TVAR parameter $\alpha(z^{\ast})=0.199$. {\em Dashed lines:} these
are the normal density functions with mean and variance equal to the sample mean and variance of the
$\hat{c}$ values. {\em Vertical dotted line:} The theoretical value of the parameter.}
\end{figure}

\begin{figure}
\centering
\resizebox{\textwidth}{!}{\includegraphics{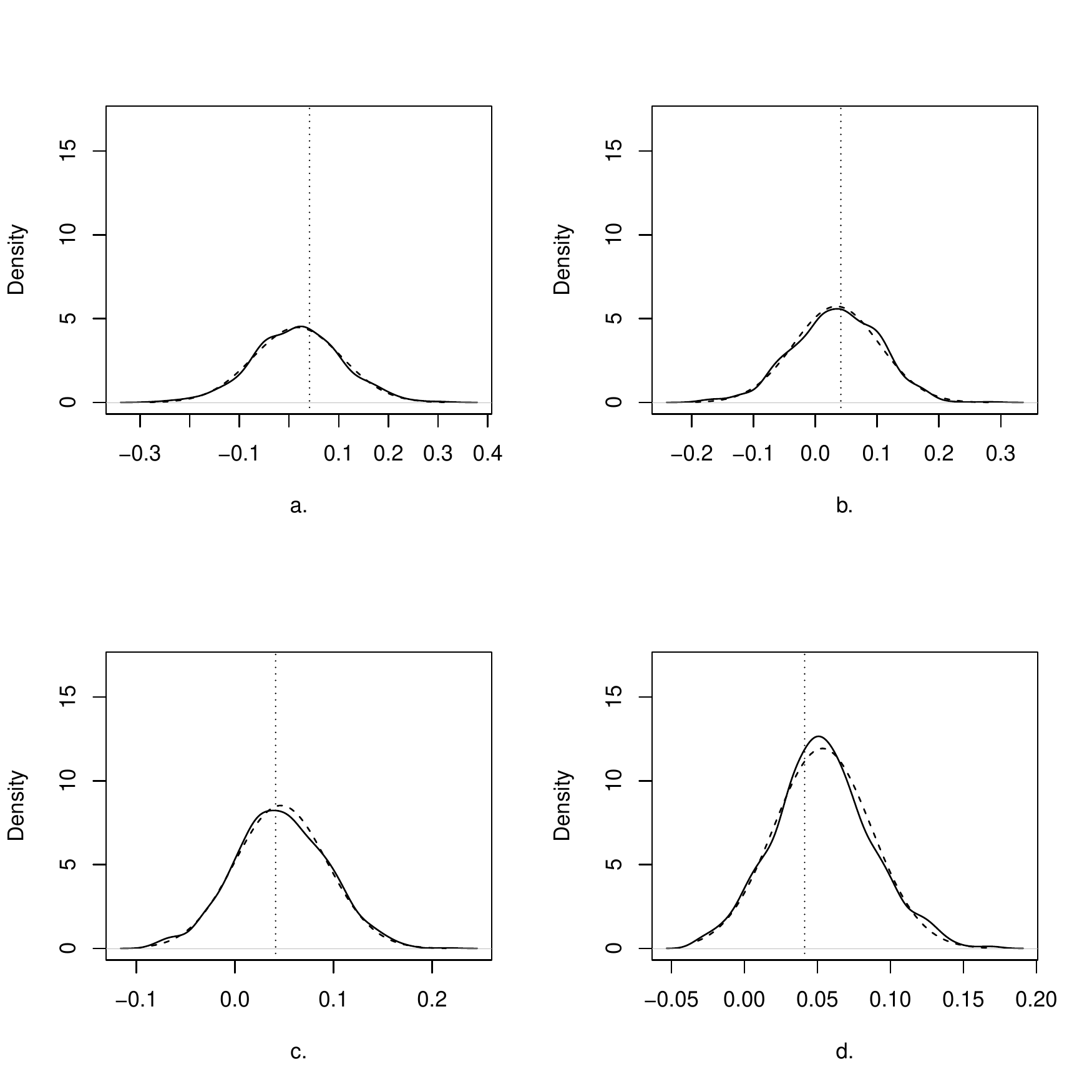}}
\caption{\label{fig:asynTVAR2} {\em Solid line:} Density estimates of values of $\hat{c}(z^{\ast}=200/512, 2)$ for
{\em a.} $T=512$, {\em b.} $T=1024$, {\em c.} $T=2048$ and {\em d.} $T=4096$ all
computed from TVAR$(1)$ model simulations with TVAR parameter $\alpha(z^{\ast})=0.199$. {\em Dashed lines:} these
are the normal density functions with mean and variance equal to the sample mean and variance of the
$\hat{c}$ values. {\em Vertical dotted line:} The theoretical value of the parameter.}
\end{figure}

\begin{figure}
\centering
\resizebox{\textwidth}{!}{\includegraphics{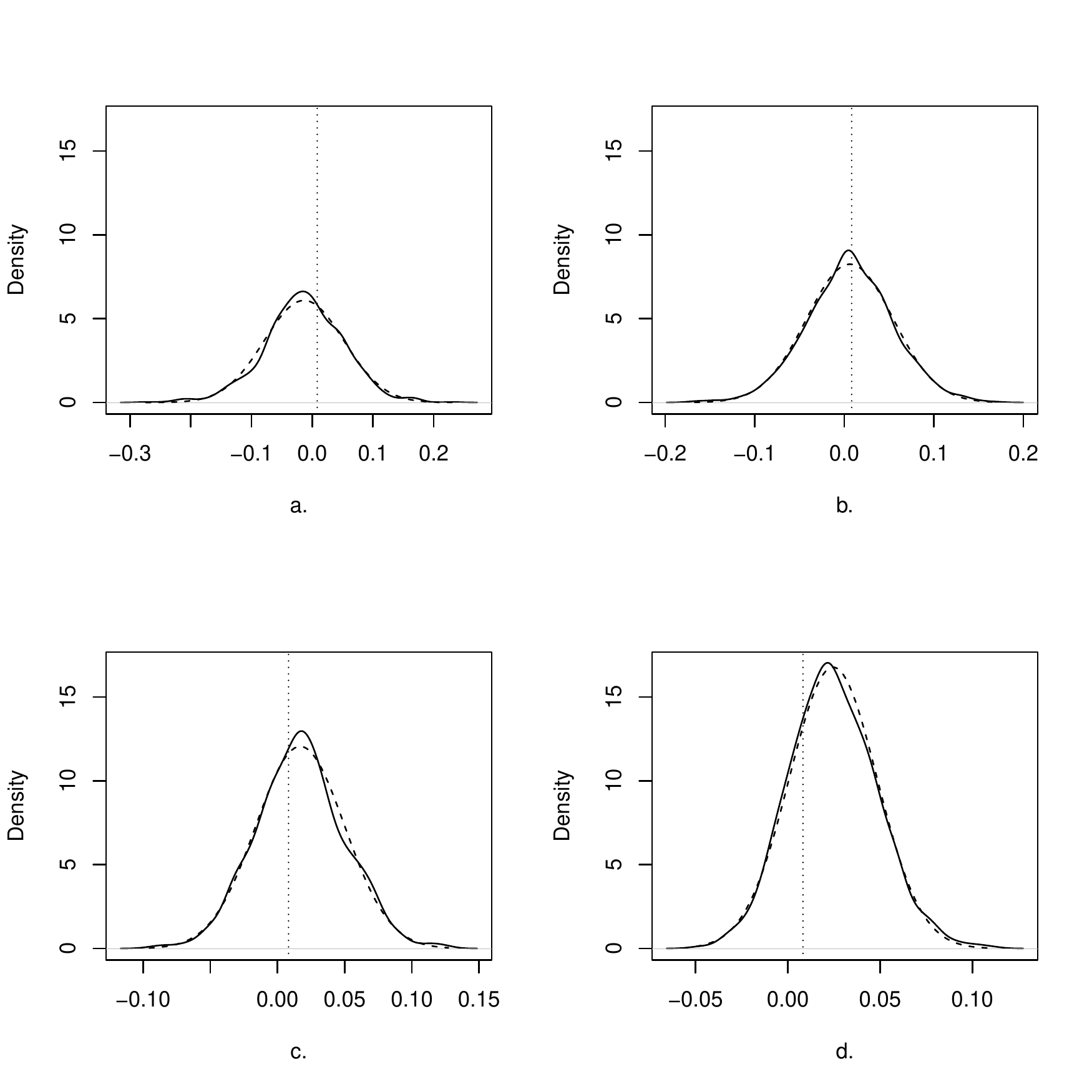}}
\caption{\label{fig:asynTVAR3} {\em Solid line:} Density estimates of values of $\hat{c}(z^{\ast}=200/512, 3)$ for
{\em a.} $T=512$, {\em b.} $T=1024$, {\em c.} $T=2048$ and {\em d.} $T=4096$ all
computed from TVAR$(1)$ model simulations with TVAR parameter $\alpha(z^{\ast})=0.199$. {\em Dashed lines:} these
are the normal density functions with mean and variance equal to the sample mean and variance of the
$\hat{c}$ values. {\em Vertical dotted line:} The theoretical value of the parameter.}
\end{figure}

Figures~\ref{fig:asynTVAR0} to~\ref{fig:asynTVAR3} provide additional evidence for the asymptotic normality
but using a nonstationary TVAR$(1)$ model selected at the particular time point $z^{\ast} = 200/512$.
The equivalent AR$(1)$ parameter at this time point is $0.199$.

All the density estimates were produced with the {\tt density()}  function in {\tt R} with default arguments.

\section{Reproducing figures in the Companion Paper}
\label{sec:rr}
In the interests of reproducible research we explain how we produce the figures in the companion paper,
Nason (2013)

\subsection{Figure 1: Earthquake P data}
The data for this section are described in Shumway and Stoffer (2006).
The earthquake and explosion data can be obtained from David Stoffer's website at
\begin{verbatim}
http://www.stat.pitt.edu/stoffer/tsa.html
\end{verbatim}
The data is stored in a file \verb+eq5exp6.dat+ as a single vector of 4096 observations. The earthquake P and Q waves are
stored first, then the explosion P and Q waves. We've stored them as \verb+eqP, eqQ, exP+ and \verb+exQ+ each vectors
of 1024 values. With this information Figure~1 in the main paper can be produced using the following commands:
\begin{verbatim}
eqP.hwtos2 <- hwtos2(eqP)
plot(eqP.hwtos2)
\end{verbatim}

\subsection{Figure 2: Explosion P data}
Similarly, the explosion P data test of stationarity plot can be obtained by
\begin{verbatim}
exP.hwtos2 <- hwtos2(exP)
plot(exP.hwtos2)
\end{verbatim}

\subsection{Figure 3: dBabyECG data}
The \verb+dBabyECG+ data is constructed by
\begin{verbatim}
dBabyECG <- diff(c(BabyECG[2], BabyECG))
\end{verbatim}
Note: the book is actually incorrect (typos), but a correct version is on the errata list of the book.
The commands to produce Figure 3 are:
\begin{verbatim}
dBabyECG.hwtos2 <- hwtos2(dBabyECG)
plot(dBabyECG.hwtos2)
\end{verbatim}

\subsection{Figure 4: various acfs of EqP data}
Figure~4 produced by the following bespoke set of commands:
\begin{verbatim}
tmp <- acf(eqP[100:200], plot = FALSE, lag.max = 30)$acf
    tmp2 <- acf(eqP[1:300], plot = FALSE, lag.max = 30)$acf
    tmp <- tmp[, , 1]
    tmp2 <- tmp2[, , 1]
    eqP.lacvCI <- Rvarlacv(eqP, nz=150)
    plot(eqP.lacvCI, type = "acf", main = "", segandcross=FALSE, sub="")
    symbols(0:30, tmp, circles = rep(0.2, 31), inches = FALSE,
        fg = 1, add = TRUE)
    sm1 <- rep(0.3, length(tmp2))
    sm2 <- rep(0.3, length(tmp2))
    sm <- cbind(sm1, sm2, sm1, sm2, sm1, sm2)
    n <- 18
    points(0:30, tmp2, pch = n)
    lin1 <- 2/sqrt(100)
    abline(h = lin1, lty = 2)
    lin2 <- 2/sqrt(300)
    abline(h = lin2, lty = 3)
\end{verbatim}

\subsection{Figure 5: Localized ACV of EqQ at two points}
Figure 5 was produced by:
\begin{verbatim}
exQ.lacv.50 <- Rvarlacv(x=exQ, nz=50, var.lag.max=30)
exQ.lacv.900 <- Rvarlacv(x=exQ, nz=900, var.lag.max=30)
plot(exQ.lacv.50, plotcor=FALSE, type="acf")
plot(exQ.lacv.900, plotcor=FALSE, type="acf")
\end{verbatim}

\subsection{Figure 6: Localized ACV of TVAR(1) process}
The four figures in Figure 6 were produced by
\begin{verbatim}
#
# Plot a.
#
x <- tvar1sim()
x.100.tvar1 <- Rvarlacv(x, nz=100, var.lag.max=25)
plot(x.100.tvar1, plotcor=FALSE, type="acf", sub="a.")
#
# Plot b.
#
x.200.tvar1 <- Rvarlacv(x, nz=200, var.lag.max=25)
plot(x.200.tvar1, plotcor=FALSE, type="acf", sub="b.")
#
# Plot c.
#
x.300.tvar1 <- Rvarlacv(x, nz=300, var.lag.max=25)
plot(x.300.tvar1, plotcor=FALSE, type="acf", sub="c.")
#
# Plot d.
#
x.400.tvar1 <- Rvarlacv(x, nz=400, var.lag.max=25)
plot(x.400.tvar1, plotcor=FALSE, type="acf", sub="d."
\end{verbatim}
Of course, the figures will be different as the realisations
from the {\tt tvar1sim()} will be different on different machines and
for different invocations.

\section*{References}

Nason, G.P. (2013) A test for second-order stationarity and approximate confidence intervals for
	localized autocovariances for locally stationary time series.
	{\em J.\ R.\ Statist.\ Soc.} B, {\bf 75}, 879--904.
	
\vspace{0.4cm}

\noindent
Priestley, M.B.\ and Subba~Rao, T. (1969) A test for non-stationarity of time series.
{\em J.\ R.\ Statist.\ Soc.} B, {\bf 31}, 140--149.

\vspace{0.4cm}

\noindent
Shumway, R.H.\ and Stoffer, D.S. (2006) {\em Time Series Analysis and Its Applications with R Examples},
	Springer: New York.

\end{document}